\documentstyle[preprint,aps]{revtex}
\widetext
\draft
\tighten

\begin{document}

%\title{\begin{center}{Electronic States}\end{center}
%\begin{center}{in Diffused Quantum Wells}\end{center}}

\title{Electronic States in Diffused Quantum Wells}

\author{{\bf S. Vlaev${^*}$,}\,\,\, {\bf D.A.  Contreras-Solorio}}
\address{Escuela de F\'{\i}sica, Universidad Aut\'onoma de Zacatecas,\\
Apartado Postal C-580,\, Zacatecas 98068, Zac., M\'exico}

\bigskip

\date{\today}

\maketitle

%\begin{center}
%(28 January, 1997)
%\end{center}

\bigskip

%\baselineskip 5.25mm

%\bigskip
%\bigskip

\begin{abstract}

In the present study we calculate the energy values and the spatial
distributions of the bound electronic states in some diffused quantum wells.
The calculations are performed within the virtual
crystal approximation, $sp^{3}s^{*}$ spin dependent empirical tight-binding
model and the surface Green function matching method. A good
agreement is found between our results and experimental data obtained for
$AlGaAs/GaAs$ quantum wells with thermally induced changes in the profile
at the interfaces. Our calculations show that for diffusion lengths
$L_{D}$=$20\div100$ {\AA} the transition $(C3-HH3)$ is not sensitive to the
diffusion length, but the transitions $(C1-HH1)$,  $(C1-LH1)$,  $(C2-HH2)$
and $(C2-LH2)$ display large "blue shifts" as $L_{D}$ increases. For
diffusion lengths $L_{D}$=$0\div20$ {\AA} the transitions $(C1-HH1)$ and
$(C1-LH1)$ are less sensitive to the $L_{D}$ changes than the $(C3-HH3)$
transition. The observed dependence is explained in terms of the
bound states spatial distributions.
\end{abstract}


\newpage
%\begin{center}
\section{Introduction}
%\end{center}

The $Al_{x}Ga_{1-x}As$/$GaAs$ heterostructures have non-abrupt
interfaces due to unwanted diffusion of $Al$ and $Ga$ across the
heterojunctions. These compositionally graded interfaces change the
electronic and optical properties of the quantum well structures [1-11]. A
detailed study about the diffusion influence on the quantum well electronic
states allows to control and use the new features these systems
display. A comparison between calculated and measured transition energies
permits to find diffusion coefficients for different temperatures
[2,4,7,8].  Till now quantum wells with compositional grading at the interfaces
were studied mainly using effective mass models and the envelope
function approximation [1-4,6,8,9,11].  We believe that realistic
tight-binding calculations will give a new information for these systems as
was illustrated in [5] for an $AlAs$/$GaAs$ quantum well with a linear
concentration profile at the interfaces.

In the present paper we study the influence of the diffusion length $L_{D}$
on the optical transitions energies in an interdiffused
$Al_{0.2}Ga_{0.8}As$/$GaAs$ quantum well with an as-grown well width of
100 {\AA}. This well was fabricated and experimentally studied in Ref.[1].
The energies of the transitions $(C1-HH1)$,  $(C1-LH1)$  and
$(C2-HH2)$  were measured at temperature 2 K for
$L_{D}$=0, 18, 20, 28, 48.5 and 66 {\AA}\,\, after annealing at 950 C
for different time durations.

%\begin{center}
\section{Model and Method}
%\end{center}

The data of the system under study are taken from Ref.[1]. The width of the
initial (as-grown) $Al_{0.2}Ga_{0.8}As$/$GaAs$ rectangular quantum well is 100
{\AA}. The $Al$ composition across the well after the diffusion can be
found from the expression: (See fig.1.)
\begin{equation} C(z) = C_b + (C_w-C_b)/2\,\left[{\tt
erf}\,(h-z)/L_D+{\tt erf}\,(h+z)/L_D\right],
\end{equation} where $C(z)$ is the $Al$ concentration at a distance $z$
from the well center; $C_w$, the $Al$ concentration in the well before the
diffusion; $C_b$, the $Al$ concentration in the barrier before the
diffusion; $2h$, the width of the initial well; $L_{D}$, the diffusion
length. The Eq.(1) is widely used in the literature (see for instance
Ref.[1]).

We use the $sp^{3}s^{*}$ spin dependent empirical tight binding model,
the virtual crystal approximation and the surface Green function
matching technique [12-15]. For more details see [12] and
references therein. The calculations were performed at the center of the
two-dimensional Brillouin zone for the (100)
growth direction.

Each diffused structure was divided into three regions, namely, the
external homogeneous barriers with negligible diffusion (the
$Al$ concentration in the planes of matching was 0.199) and the internal
inhomogeneous (diffused) well region having a concentration profile as
determined from Eq.(1). The size of the inhomogeneous slab was 167, 131,
and 95 monolayers respectively for $L_{D}$=100, 70 and 35 {\AA}. The Green
function of the external barriers was calculated from the transfer matrix
in the usual way and the Green function of the diffused well region,
by means of the algorithm established and used to study other
heterogeneous structures [12,13,15].

%\begin{center} \section{Results and Discussion} %\end{center}
\section{Results and Discussion}

We calculated the energies of the first three bound electron states $C1$,
$C2$, $C3$ (see fig.2a) and the first five hole states $HH1$, $LH1$, $HH2$,
$HH3$ and $LH2$ (see fig.2b) in the studied quantum well for different
diffusion lengths $L_{D}$. The calculations in the interval $L_{D}=0\div70$
{\AA} were performed through 5 {\AA} and in the interval $L_{D}=70\div100$
{\AA} - through 10 {\AA}. We also found the bound states energies for the
diffusion lengts $L_{D}$=0, 18, 20, 28, 48.5 and 66 {\AA}. (The
measurements in [1] were conducted with these diffusion lengths values.)
As $L_{D}$ increases the bound states depart further from the $GaAs$
band edges.  Qualitatively similar results have been reported in [5] for an
$AlAs$/$GaAs$ quantum well with linearly graded interfaces. As a difference
from [5], in our case the energies of the states $C3, HH3$  and $LH2$
change several times less than the energies of the states $C1, C2, HH1,
HH2$  and  $LH1$. For instance, over the whole interval $L_{D}=0\div70$
{\AA} the state $C3$ rises only by 8 meV while over the same interval the
state $C1$ rises by 61 meV. The states $HH3$ and $HH1$ change by 19 meV and
36 meV respectively over the same diffusion length interval.

To understand the above mentioned behaviour we calculated the spatial
distributions of all bound states for all diffusion lengths.
The diffusion modifies the initial (as-grown) quantum well and leads
to formation of two
regions in the diffused well. In the as-grown well region (see fig.1) the
$Al$ concentration increases and the energies of the bound
electron (hole) states
in the diffused quantum well increase (decrease). (The energy
zero is at the $AlAs$ valence band top edge.) On the contrary, in the
as-grown barrier region (see fig.1) the $Al$ concentration decreases
and the energies of the bound states decrease (increase) for electrons
(holes). All bound states are more concentrated in the as-grown
well region and consequently their energies increase (decrease) for
electrons (holes) as $L_{D}$ increases.
But, in general, a competition exists between two regions
affected by the diffusion and a compensation of the energy shifts
can occur. The magnitude of the energy
shift for a given bound state depends on the relation between the
probabilities of this state to occupy the diffused well part with increased
$Al$ concentration (the as-grown well region) and with decreased $Al$
concentration (the as-grown barrier region). As an example, in order to
illustrate the role of the above mentioned compensation, the spatial
distributions of the electron states $C3$ and $C1$ are shown in fig.3a and
fig.3b for the as-grown and a diffused ($L_{D}$=35{\AA}) quantum
well. The energy of the state $C3$ changes only by 7 meV because this
state spreads over both
regions affected by the diffusion (within and outside the
initial well, see fig.3a) and
the compensation occurs. The hole state $HH3$ has a similar
spatial distribution and also shifts weakly. The
energy of the state $C1$ shifts with 21 meV due to its spatial
localization in the
central well region with increased $Al$ concentration
after the diffusion (see fig.3b) and
the compensation does not occur. The same argument is valid for the hole
states $HH1$ and $LH1$.

Theoretical curves of the transition energies
dependence on the diffusion length $L_{D}$ for the transitions $(C1-HH1)$,
$(C1-LH1)$, $(C2-HH2)$, $(C2-LH2)$  and  $(C3-HH3)$ are shown in fig.4a for
$Al_{0.2}Ga_{0.8}As$/$GaAs$ diffused quantum wells. There is a "blue shift"
for all transitions with $L_{D}$ increasing as reported in [2,4] for
similar systems. This shift is 27 meV for $(C3-HH3)$  and  97 meV for
$(C1-HH1)$ as $L_{D}$ increases from 0 to 70 {\AA}. The transitions
$(C1-LH1)$, $(C2-HH2)$  and  $(C2-LH2)$ shift by 94 meV, 82 meV and 65 meV
respectively.

The transition $(C3-HH3)$ is not sensitive to
the diffusion length for $L_{D}$=$20\div100$ {\AA} because the energies of
the states $C3$ and $HH3$ change weakly, see fig.2. A relatively low
sensitivity of the transition $(C3-HH3)$ to the diffusion length was
measured in [2].
The transitions $(C1-HH1)$, $(C1-LH1)$  and
$(C2-HH2)$ show considerable "blue shifts" in the interval
$L_{D}$=$20\div100$ {\AA} because the energies of the states $C1$, $C2$,
$HH1$, $LH1$ and $HH2$ change significantly, see fig.2.

For low diffusion length values ($L_{D}$ $\leq20$ {\AA}) the transition
$(C3-HH3)$ is sensitive to the $L_{D}$ changes because the energies of the
states $C3$ and $HH3$ depend strongly on the diffusion length, see fig.2.
The $Al$ concentration changes only near the interfaces and the spatial
distributions of the states $C3$ and $HH3$ do not allow a compensation.
In this diffusion length interval the transitions $(C1-HH1)$ and
$(C1-LH1)$ are less sensitive to the $L_{D}$ changes due
to the energies of the states $C1$, $HH1$ and $LH1$ depend on $L_{D}$
weaker than the $C3$ and $HH3$ energies, see
fig.2. The states $C1$, $HH1$ and $LH1$ are mainly localized in the
central well part where the $Al$ concentration is not affected.

In fig.4b we compare our theoretical curves with the experimental curves
from  [1]  for $(C1-HH1)$, $(C1-LH1)$ and $(C2-HH2)$ transitions energies
with $L_{D}$=0, 18, 20, 28, 48.5 and 66 {\AA}. The
measurements in [1] have been conducted at temperature 2 K and
temperature corrections of the theoretical values were not necessary. It is
obvious that the agreement with the experimental data is quite satisfactory
inspite of the fact that excitonic effects were not taken into account and
fit was not done.
All experimental curves lie about $10\div45$ meV below the
theoretical ones. The agreement between the theory
and the experiment is better for the excited states than the ground states
due to the fact that the exciton binding energies are higher for the ground
states. For all transitions in fig.4b two curves (a theoretical and the
experimental) diverge with $L_{D}$
increasing. For instance, for $L_{D}$=0 {\AA} the discrepances are 29 meV,
30 meV and 10 meV for the transitions $(C1-HH1)$, $(C1-LH1)$ and $(C2-HH2)$,
respectively, and for $L_{D}$=66 {\AA} 45 meV, 41 meV and 29 meV respectively.
A possible explanation of this behaviour could be an increase of the
excitonic binding energy for higher $L_{D}$ values due to the higher $Al$
concentration within the initial well region. This means that the excitons
are localized even in the case of $L_{D}$=66 {\AA}. The
spatial distributions of the bound states for these diffusion lengths
support such an assumption. An exciton binding energy increase has been
reported in [4] for double-barrier quantum wells as $L_{D}$ increases.

\newpage
%\begin{center}
\section{Conclusion}
%\end{center}

In the present paper we performed numerical calculations which showed that
the empirical tight-binding model combined with the surface Green function
matching method and the algorithm for treating inhomogeneous finite
slabs [12-15] works very well in case of quantum wells with graded
interfaces. We found that the theoretical dependence of the transition
energies on the diffusion lengths is in good agreement with the experiment.

For given as-grown well width and diffusion length $L_{D}$, the sensitivity
of the transitions to $L_{D}$ changes depends on the bound states
spatial distributions over the diffused well regions. For small $L_{D}$
values (in our case below 20 {\AA}) the transition $(C3-HH3)$ is more
sensitive to the $L_{D}$ changes than the transitions $(C1-HH1)$ and
$(C1-LH1)$. For large $L_{D}$ values (in our case above 20 {\AA}) the
transition $(C3-HH3)$ is not sensitive to the diffusion length, but the
transitions $(C1-HH1)$ and $(C1-LH1)$ display large "blue shifts".

\bigskip
\noindent
-----------------------------------------------------------------
\newline
*Permanent address: Institute of General and Inorganic Chemistry,
Bulgarian Academy of Sciences, 1113 Sofia, Bulgaria
%\newpage
%\medskip
%\begin{center}
\acknowledgments
%\end{center}

We are grateful to Prof.V.Velasco, Prof.F.Garc\'{\i}a-Moliner and
Prof.A.Chubykalo for helpful discussions. SV is specially indebted to the
CONACyT, M\'exico for support. This work was partially supported by the
National Fund for Scientific Investigations, Bulgaria through Grant X-646,
CONACyT (M\'exico), through grant 1852p-E, and European
Community, through grant CI1*-CT94-006

\newpage
\begin{figure}
\caption{Al concentration profiles of quantum wells for
diffusion lengths $L_{D}$=0, 25, 50 and 100 {\AA}. The value of $L_{D}$=0
{\AA} corresponds to the initial as-grown rectangular quantum well
of 100 {\AA} (35 monolayers).}
\end{figure}
\begin{figure}
\caption{(a) Electron energy levels (in eV) of the states $C1$, $C2$ and
$C3$, (b) hole energy levels (in eV) of the states $HH1$, $LH1$, $HH2$,
$HH3$ and $LH2$ for diffused quantum wells, as a function of the diffusion
length $L_{D}$ (in {\AA}).  The value of $L_{D}$=0 {\AA} corresponds to the
well width of 100 {\AA} (35 monolayers) of the initial as-grown rectangular
quantum well. The energy zero is at the $AlAs$ valence band top edge.}
\end{figure}

\begin{figure}
\caption{Spatial distributions of $C3$ (a) and $C1$ (b)
electron bound states over the as-grown (solid lines) and diffused (dotted
lines) quantum wells. The diffusion length is $L_{D}$=35 {\AA}.
The bound states energies are: $C3$ (as-grown well) - 2.240 eV,
$C3$ (diffused well) - 2.247 eV;   $C1$ (as-grown well) - 2.118 eV,
$C1$ (diffused well) - 2.139 eV.}
\end{figure}

\begin{figure}
\caption{
Transition energies (in eV) for diffused quantum wells, as a function of
the diffusion length $L_{D}$ (in {\AA}).  The value of $L_{D}$=0 {\AA}
corresponds to the well width of 100 {\AA} (35 monolayers) of the initial
as-grown rectangular quantum well.
(a) theory; transitions $(C1-HH1)$,
$(C1-LH1)$, $(C2-HH2)$, $(C2-LH2)$ and $(C3-HH3)$.
(b) theory (th.) and experiment
(exp.); transitions $(C1-HH1)$, $(C1-LH1)$ and $(C2-HH2)$.}
\end{figure}

\end{document}